\documentclass[aps, reprint, superscriptaddress,prl]{revtex4-2}
\usepackage{graphicx}
\usepackage{placeins}
\usepackage{textcomp}
\usepackage{amssymb}
\usepackage{amsmath}
\usepackage[T1]{fontenc}
\usepackage{times, txfonts}
\usepackage{makecell}
\usepackage{physics}
\usepackage[english]{babel}
\usepackage[colorlinks=true,linkcolor=blue,urlcolor=blue,citecolor=blue,pdfusetitle]{hyperref}

\usepackage{enumitem}
\usepackage{xcolor}
\definecolor{c_XX}{RGB}{190, 24, 26}
\definecolor{c_X}{RGB}{76, 110, 180}

\begin{document}

\title{Indistinguishable photons from a two-photon cascade}

\author{Timon L. Baltisberger}
\thanks{These authors contributed equally}
\affiliation{Department of Physics, University of Basel, Klingelbergstrasse 82, 4056 Basel, Switzerland}

\author{Francesco Salusti}
\thanks{These authors contributed equally}
\affiliation{Department of Physics and Center for Optoelectronics and Photonics Paderborn (CeOPP), Paderborn University, Warburger Strasse 100, 33098 Paderborn, Germany}
\affiliation{Institute for Photonic Quantum Systems (PhoQS), Paderborn University, 33098 Paderborn, Germany}

\author{Mark R. Hogg}
\affiliation{Department of Physics, University of Basel, Klingelbergstrasse 82, 4056 Basel, Switzerland}

\author{Malwina A. Marczak}
\affiliation{Department of Physics, University of Basel, Klingelbergstrasse 82, 4056 Basel, Switzerland}

\author{Nils Heinisch}
\affiliation{Department of Physics and Center for Optoelectronics and Photonics Paderborn (CeOPP), Paderborn University, Warburger Strasse 100, 33098 Paderborn, Germany}
\affiliation{Institute for Photonic Quantum Systems (PhoQS), Paderborn University, 33098 Paderborn, Germany}

\author{Sascha R. Valentin}
\affiliation{Faculty of Physics and Astronomy, Experimental Physics VI, Ruhr University Bochum, Universitätsstrasse 150, 44801 Bochum, Germany}

\author{Stefan Schumacher}
\affiliation{Department of Physics and Center for Optoelectronics and Photonics Paderborn (CeOPP), Paderborn University, Warburger Strasse 100, 33098 Paderborn, Germany}
\affiliation{Institute for Photonic Quantum Systems (PhoQS), Paderborn University, 33098 Paderborn, Germany}
\affiliation{Wyant College of Optical Sciences, University of Arizona, Tucson, AZ 85721, USA}

\author{Arne Ludwig}
\affiliation{Faculty of Physics and Astronomy, Experimental Physics VI, Ruhr University Bochum, Universitätsstrasse 150, 44801 Bochum, Germany}

\author{Klaus D. J{\"o}ns}
\affiliation{Department of Physics and Center for Optoelectronics and Photonics Paderborn (CeOPP), Paderborn University, Warburger Strasse 100, 33098 Paderborn, Germany}
\affiliation{Institute for Photonic Quantum Systems (PhoQS), Paderborn University, 33098 Paderborn, Germany}

\author{Richard J. Warburton}
\affiliation{Department of Physics, University of Basel, Klingelbergstrasse 82, 4056 Basel, Switzerland}


\begin{abstract}
Decay of a four-level diamond scheme via a cascade is a potential source of entangled photon pairs. A solid-state implementation is the biexciton cascade in a semiconductor quantum dot. While high entanglement fidelities have been demonstrated, the two photons, XX and X, are temporally correlated, typically resulting in poor photon coherence. Here, we demonstrate a high two-photon interference visibility (a measure of the photon coherence) for both XX ($V$=94$\pm$2\%) and X ($V$=82$\pm$6\%) photons. This is achieved by Purcell-enhancing the biexciton transition in a low-noise device. We find that the photon coherence follows the well-known quantum optics result upon tuning the XX:X lifetime ratio over two orders of magnitude.
\end{abstract}

\maketitle

\newcommand{\decayXXtoX}{$\ket{\text{XX}} \hskip -0.2mm \rightarrow \hskip -0.2mm \ket{\text{X}}$}
\newcommand{\decayXtoGS}{$\ket{\text{X}} \hskip -0.2mm \rightarrow \hskip -0.2mm \ket{\text{0}}$}

Entangled photons are a key resource in quantum communication \cite{Gisin2007}. For basic quantum key distribution following the E91 protocol, a two-photon entangled state is sufficient \cite{Ekert1991}. More elaborate schemes involve entanglement swapping for which the photons in separate pairs must be indistinguishable \cite{Pan1998}. Parametric down conversion using a non-linear crystal can be used to create such pairs but such sources are probabilistic \cite{Kwiat1995, Zhong2018}. A deterministic source of indistinguishable entangled photon pairs, particularly one operating at a high repetition rate, would represent a radical improvement. An option here is to use a semiconductor quantum dot (QD). A QD has a number of attractive features: the optical dipole moments are large such that operation at GHz rates is feasible \cite{Rickert2025}; the interaction with phonons is weak such that the photons are coherent; noise can be suppressed in state-of-the-art heterostructures \cite{Kuhlmann2013}; and photonic engineering can be used to create high efficiencies \cite{somaschi2016near, wang2016near, uppu2020scalable, Tomm2021, Ding2025, liu2025quantum}. However, while creating indistinguishable single photons with a QD has been achieved, an outstanding challenge is to create highly indistinguishable, entangled photon pairs.

\begin{figure}[t!]
    \centering
    \includegraphics{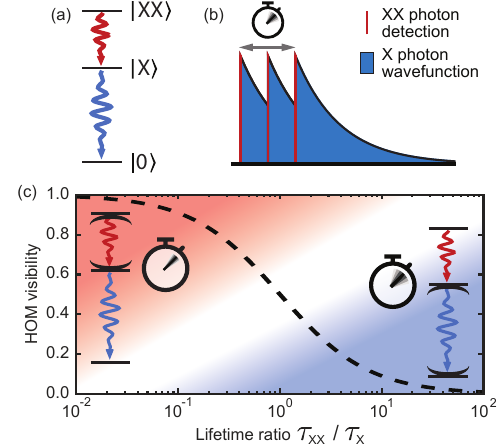}
    \caption{(a) Schematic of the biexciton cascade. The biexcition decays to the exciton (\decayXXtoX) creating photon XX; the exciton decays to the ground state (\decayXtoGS) creating photon X. Only one decay pathway is shown for simplicity. 
    (b) The XX and X photons are correlated in time limiting the indistinguishability of successive XX photons, likewise X photons. The sketch shows a simplified picture: detection of the XX results in timing jitter in the X photon.
    (c) Theoretical prediction of the two-photon interference visibility as a function of lifetime ratio, Eq.~(\ref{eq:HOMvsLifetime}). Reducing $\tau_{\text{XX}}$ by Purcell enhancing \decayXXtoX \hskip 0.8mm reduces the timing jitter and improves the coherence of both XX and X photons (red shaded area). Conversely, reducing $\tau_\text{X}$ has the opposite effect (blue shaded area).
    }
    \label{cascade}
\end{figure}

A charge-neutral QD can be doubly excited to create a biexciton, a spin-less complex of two electrons and two holes. The biexciton decays to the exciton (\decayXXtoX) creating the XX photon; the exciton decays to the ground state (\decayXtoGS), creating the X photon, Fig.~\ref{cascade}(a). Two decay paths are available opening up a scheme to create an entangled photon pair \cite{Benson2000, Akopian2006}. On eliminating the splitting between the two exciton states (the ``fine structure") \cite{Trotta2012}, which-path information is erased and the two photons are entangled in the polarisation basis \cite{Huber2018strain}. This scheme has been widely implemented and in the best case yields very high entanglement fidelities, up to 98.7\% \cite{Huber2018strain,Schimpf2021,pennacchietti2024}. Alternatively, one decay path can be used to create an entangled photon pair in the time-bin basis \cite{Simon2005}: the system should be prepared in a metastable state, perhaps a ``dark" exciton \cite{kappe2025keeping}, with an optical transition to the biexciton. The biexciton is then created using a $\frac{\pi}{2}$-pulse (``early" time-bin) followed by a $\pi$-pulse (``late" time-bin). Work is underway to implement this alternative scheme \cite{aumann2022demonstration}. With respect to photon coherence, both schemes suffer from the same drawback. Inevitably, the XX photon is detected before the X photon resulting in an unwanted temporal correlation and a loss of photon coherence (as quantified via the indistinguishability metric, the two-photon interference visibility $V$) \cite{Schoell2020}. In simple terms, timing jitter reduces the coherence of both XX and X photons, Fig.~\ref{cascade}(b).

Quantum optics (Wigner-Weisskopf theory with no biexciton or exciton dephasing) predicts that the coherence metric $V$ for both photons is given by \cite{huang1993correlations, Simon2005, Schoell2020}
\begin{equation}
V=\frac{1}{1+\tau_\text{XX}/\tau_\text{X}}.
\label{eq:HOMvsLifetime}
\end{equation}
In the standard case, $\tau_\text{XX}\simeq 0.65 \, \tau_\text{X}$ \cite{dalgarno2008coulomb}, giving $V \simeq 0.6$, a modest value. Eq.~(\ref{eq:HOMvsLifetime}), points to a powerful mitigation strategy; a microcavity can be used to manipulate $\tau_\text{XX}$ and $\tau_\text{X}$ via the Purcell effect \cite{gerard1998enhanced}, Fig.~\ref{cascade}(c). Once $\tau_\text{XX} \ll \tau_\text{X}$, $V$ is close to unity. On QDs, Purcell factors of up to 40 have been achieved \cite{Liu2018} such that $V$ is potentially as high as 98.8\%. Eq.~(\ref{eq:HOMvsLifetime}) has not been probed experimentally over a wide range of lifetime ratio $\tau_\text{XX}/\tau_\text{X}$. It is an open question to what extent Eq.~(\ref{eq:HOMvsLifetime}) applies to the biexciton cascade given that the theory neglects all the complexities of the semiconductor (biexciton/exciton phonon scattering, charge and spin noise). Previous attempts at controlling $V$ relied either on dot-to-dot fluctuations in the lifetime ratio \cite{Schoell2020} or on tuning via an applied electric field \cite{Undeutsch2025}. In the first case, $V$ was limited to a small range around 60\%. In the second case, while $V_\text{XX}$ could be increased to 76.9\%, the electric field regime required to increase $\tau_\text{X}$ and imbalance the lifetime ratio reduced $V_\text{X}$ to 42\%: the improvement in XX photon coherence thus came at the expense of X coherence, a clear departure from Eq.~(\ref{eq:HOMvsLifetime}).

Here, we explore experimentally the dependence of $V$ on lifetime ratio upon tuning $\tau_\text{XX}/\tau_\text{X}$ over two orders of magnitude. In one extreme, $\tau_\text{XX}/\tau_\text{X}=0.08$, we demonstrate $V_\text{XX}=94\pm2\%$ and $V_\text{X}=82\pm6\%$, improving the coherence for both XX and X photons. Conversely, in the other extreme, $\tau_\text{XX}/\tau_\text{X}=6.2$ , $V_\text{XX}=12\pm3\%$ ($V_\text{X}=16\pm1\%$ for $\tau_\text{XX}/\tau_\text{X}=5.5$), closely matching the theoretical prediction. 

\begin{figure}[t!]
    \centering
    \includegraphics{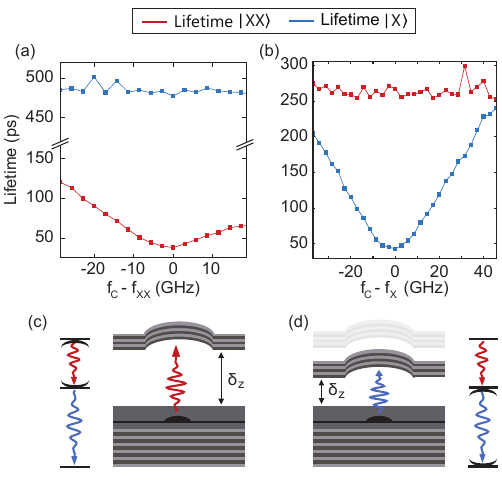}
    \caption{Transition lifetimes detecting the X and XX photons upon scanning the cavity resonance over (a) \decayXXtoX \hskip 0.8mm and (b) \decayXtoGS. $f_\text{C}$ is the cavity resonance-frequency, $f_\text{XX}$ ($f_\text{X}$) the XX (X) resonance-frequency. Changing the cavity length $\delta_z$ allows $\tau_\text{XX}/\tau_\text{X}$ to be tuned \textit{in situ} over two orders of magnitude. The asymmetry in the $\ket{\text{XX}}$ lifetime in (a) comes from enhancement of the transition by the second cavity mode at higher positive detunings. 
    (c) Schematic of the open microcavity enhancing \decayXXtoX, and
    (d) enhancing \decayXtoGS.}
    \label{lifetime}
\end{figure}

The experiment uses a self-assembled InGaAs QD in an open microcavity \cite{barbour2011tunable, Tomm2021}, Fig.~\ref{lifetime}(c,d). The bottom mirror is a high-reflectivity distributed Bragg reflector within the semiconductor heterostructure. The reflectivity of the top mirror, a curved micro-mirror fabricated in a silica substrate, is lower and is chosen to optimise the end-to-end efficiency \cite{Tomm2021}. The microcavity has finesse 500, Q-factor 10,000, and, on the charged exciton, results in a maximum Purcell factor of 10. The QD charge is controlled by Coulomb blockade using a \hbox{n-i-p} diode and is set in these experiments to zero. The biexciton is created by resonant two-photon excitation (TPE) using a \mbox{5-ps} laser pulse tuned in frequency to lie halfway between the \decayXXtoX \hskip 0.8mm and \decayXtoGS \hskip 0.8mm lines. Both XX and X photons are collected; their frequency separation (the biexciton binding energy) is 690 GHz. In the output, either XX or X photons are selected with a spectral filter, and the quantum optical properties (autocorrelation $g^{(2)}$ and Hong-Ou-Mandel $V$) are determined following standard protocols \cite{sup}. One potential complication is the exciton fine structure (2.6 GHz), likewise the mode-splitting (50 GHz) of the fundamental cavity mode \cite{Tomm2021, tomm2021tuning}. The mode-splitting is sufficiently large that if one mode is close to resonance then the other mode can be ignored. Moreover, we estimate that the exciton and cavity axes are misaligned by less than 5 degrees. This is not coincidental: both are strongly influenced by the semiconductor crystal axes \cite{tomm2021tuning}. It is therefore safe to describe the experiments in terms of one cascade pathway and one cavity mode.

There are two key enabling features in these experiments. First, the microcavity frequency can be tuned \textit{in situ}, in practice by controlling the separation $\delta z$ between the semiconductor heterostructure and the top mirror, Fig.~\ref{lifetime}(c,d). The cavity mode has a $\delta z$-linewidth of 920~pm, and the root-mean-square noise in $\delta z$ is much less, just 10~pm \cite{greuter2014small}. This allows us to tune precisely the cavity mode into resonance either with \decayXXtoX \hskip 0.8mm or \decayXtoGS \hskip 0.8mm all on the same QD, hence to explore $V$ over a large range of lifetime ratios. Second, the heterostructure operates with very little charge noise \cite{Kuhlmann2013, Tomm2021}. The transition linewidths are in all cases close to the transform limit and much smaller than the cavity linewidths. This means that the Purcell factors are not subject to noise. Also, the noise at frequencies above 100~kHz is almost non-existent such that the HOM experiments, carried out here by interfering photons separated in time by 13.1 ns, are unaffected by the residual weak and slow noise processes.

We first demonstrate control of the $\ket{\text{XX}}$ and $\ket{\text{X}}$ lifetimes via the Purcell effect. When \decayXXtoX \hskip 0.8mm (\decayXtoGS) is in resonance with the microcavity mode we collect XX (X) photons efficiently and X (XX) photons very inefficiently. In the on-resonance case, the photons are emitted preferentially into the microcavity mode from which they escape through the top mirror and propagate to the detector. In the out-of-resonance case, the photons are emitted preferentially into other modes. Nevertheless, a residual coupling to the microcavity occurs via the frequency tails of the microcavity mode despite the large difference between detuning (690~GHz) and microcavity linewidth (25~GHz) \cite{Hogg2025}. This enables us to detect both XX and X photons for all microcavity detunings \cite{sup}. Plotted in Fig.~\ref{lifetime}(a,b) are the measured $\ket{\text{XX}}$ and $\ket{\text{X}}$ lifetimes as a function of microcavity detuning \cite{sup}. There is a clear Purcell effect on both $\ket{\text{XX}}$ and $\ket{\text{X}}$. The Purcell factors are 6.9 and 11.2, respectively. The difference in Purcell factors originates from the two decay pathways for the $\ket{\text{XX}}$ compared to the single pathway for each of the $\ket{\text{X}}$ \cite{sup}.

\begin{figure}[t!]
    \centering
    \includegraphics{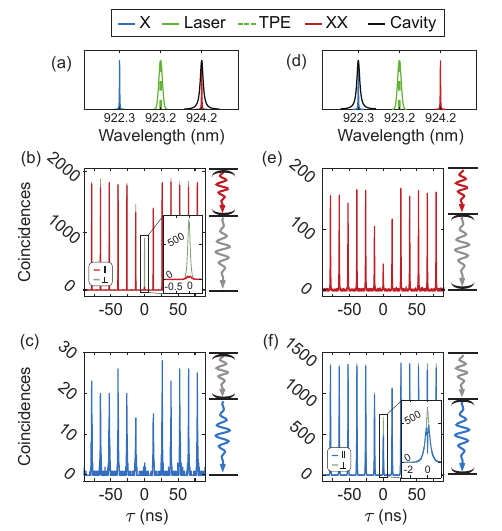}
    \caption{Characterisation of the photon coherence using two-photon interference. 
    (a) Frequency alignment of the X, XX, TPE laser and cavity for Purcell enhancement of the XX transition to reduce $\tau_{XX}$. This configuration improves the photon coherence for both XX and X photons.
    (b) Two-photon interference data for the XX photons, with $V_{XX}=94\pm2\%$. 
    (c) Two-photon interference data for the X photons, with $V_{X}=82\pm6\%$.
    (d) Frequency alignment for Purcell enhancement of the X transition, which increases $\tau_{XX}/\tau_{X}$ and thus reduces the photon coherence for both XX (e) and X (f) photons.
    }
    \label{fig:HOM_data}
\end{figure}

We proceed to measure $g^{(2)}(0)$ and $V$ for both XX and X as a function of microcavity detuning. The experiments are straightforward on the transition resonant with the microcavity. In addition, even these two-photon correlations can be performed out-of-resonance with the microcavity. However, the out-of-resonance fluxes are low and long integration times are needed, therefore we sample the theoretical prediction only at the two extreme points (XX when X is resonant with the microcavity, and vice versa). Both $g^{(2)}(0)$ and HOM are determined by adding up all the counts in a 13.1-ns window around delay zero (13.1~ns is the temporal separation of the laser pulses) \cite{sup}. Results are shown in Fig.~\ref{fig:HOM_data}. When XX (X) is in resonance with the microcavity, we measure $g^{(2)}(0)=2.3\pm0.2$\% ($g^{(2)}(0)=0.8\pm0.1$\%). We observe a slight dependence of $g^{(2)}(0)$ on the cavity detuning \cite{sup}. The finite $g^{(2)}(0)$ is likely caused by laser leakage into the detection channel (imperfect polarisation and spectral filtering) and phonon-assisted cavity feeding \cite{theory_paper}. All g$^{(2)}$ data is shown in the supplementary materials \cite{sup}. The non-zero $g^{(2)}(0)$ has an impact on the measured HOM visibility $V_{\rm raw}$. We estimate $V$ in the limit that the laser leakage is eliminated via $V=(1+2 g^{(2)}(0)) \hskip 0.5mm V_{\rm raw}$, a result derived with the assumption that in a two-photon event (X or XX plus laser photon) the two photons are completely distinguishable.

Fig.~\ref{fig:HOM_data} shows our central result. When the microcavity is resonant with \decayXXtoX, both XX and X photons have a high value of $V$, 94$\pm2$\% and 82$\pm6$\%, respectively. Conversely, when the microcavity is resonant with \decayXtoGS, both XX and X photons have a low value of $V$, 12$\pm3$\% and 16$\pm1$\%, respectively. In the former case, $\ket{\text{XX}}$ decays much more rapidly than $\ket{\text{X}}$, such that the timing jitter of X becomes less important. Conversely, in the latter case, $\ket{\text{XX}}$ decays much more slowly than $\ket{\text{X}}$ such that the timing jitter of X becomes more important \cite{theory_paper}.

\begin{figure}[t!]
    \centering
    \includegraphics{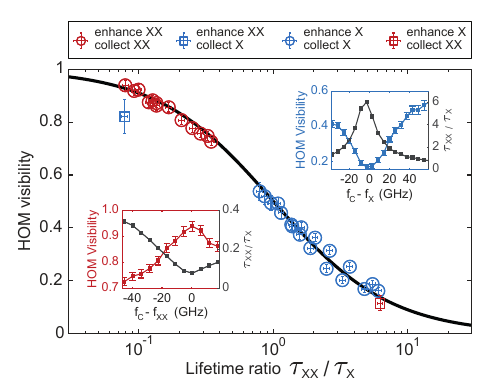}
    \caption{Photon coherence $V$ as a function of lifetime ratio $\tau_{XX}/\tau_{X}$. The solid line is Eq.~(\ref{eq:HOMvsLifetime}). Insets: HOM visibility and lifetime ratio as a function of cavity detuning from XX (red) and X (blue) transitions.}
    \label{V}
\end{figure}

We proceed to plot $V$ as a function of the lifetime ratio in Fig.~\ref{V}. The figure uses all the data, i.e., both XX and X for all cavity detunings. Within error bars, the on-resonance and out-of-resonance points lie on the same curve, evidence for a general behaviour, as predicted by Eq.~(\ref{eq:HOMvsLifetime}). To describe the system theoretically, we adapt the known theory \cite{huang1993correlations} to our case. First, photon emission takes place either into the cavity and via cavity leakage, into the one-dimensional (1D) continuum of the collection fibre, or, alternatively, into a mode confined at the semiconductor surface, also a 1D continuum. We therefore sum over 1D continua in the Wigner-Weisskopf approach. Second, we treat the QD-cavity system as a giant atom neglecting the possibility of photon reabsorption by the QD, the ``fast"-cavity limit. The result is exactly Eq.~(\ref{eq:HOMvsLifetime}). We therefore plot Eq.~(\ref{eq:HOMvsLifetime}) in Fig.~\ref{V}. We find excellent agreement with the experimental results. The theory takes into account only the temporal correlations of the XX and X photons. Dephasing via phonons, likewise charge and spin noise, are neglected.

To conclude, we state that we have manipulated the coherence of the two photons in a two-photon cascade. When the upper transition (here the biexciton in a QD) is Purcell-enhanced by a microcavity, both photons become highly indistinguishable. Conversely, when the lower transition (here the exciton) is Purcell-enhanced, both photons become distinguishable. The essential quantum-optics result describes the experiments over two decades in lifetime ratio.

Our work demonstrates that the biexciton cascade is capable of producing photons with a high degree of indistinguishability. To exploit this in a source of entangled photons in the polarisation basis, it is necessary that precession caused by the fine structure splitting is much slower than $\tau_{X}$ and there are established methods to do this \cite{Huber2018strain}.  In the time-bin basis, it is necessary to find a metastable state with an optical transition to the biexciton. In both schemes, it is also necessary to extract both XX and X photons efficiently. The open microcavity is a narrowband device and is not immediately suited to this task: if the XX photon  enters the cavity mode, the X photon couples predominantly to the surface mode. In this case, the X photons could be coupled out of the chip using an on-chip grating \cite{zhou2018high}. A simpler approach is to use a bulls-eye design for which the Purcell enhancement is wavelength-selective yet the diffraction efficiency is high over a much higher spectral range \cite{liu2019solid, bauch2024demand}. Provided the bulls-eye retains the low noise of the starting material and that the Purcell peak can be matched in resonance with the biexciton, our work shows that the biexciton decay will result in indistinguishability metrics above 90\%.

The work in Basel was funded by Swiss National Science Foundation project 200020\_204069. 
The work in Paderborn is supported by the Deutsche Forschungsgemeinschaft (German Research Foundation) through the transregional collaborative research center TRR142/3-2022 (231447078) and by the German Federal Ministry of Research,
Technology and Space (BMFTR) through the project QR.N (16KIS2206). This project
has received funding from the European Research Council (ERC) under the European Union’s Horizon 2020 research and innovation program (LiNQs, grant agreement
101042672). 
K.J. and S.S. acknowledge funding from the Ministry of Culture and Science of North Rhine-Westphalia for the Institute for Photonic Quantum Systems (PhoQS). 
S.R.V. and A.L. gratefully acknowledge funding by the EUROSTARS project QTRAIN under BMFTR grant 13N17328, the QuantERA project under BMFTR grant 16KIS2061, and QR.N under BMFTR grant 16KIS2200.

\bibliography{paper_references.bib}

\onecolumngrid
\newpage
\appendix
\setcounter{figure}{0}
\renewcommand{\thefigure}{S\arabic{figure}}

\section*{1. Experimental Set-up}

\begin{figure}[h!]
    \centering
    \includegraphics{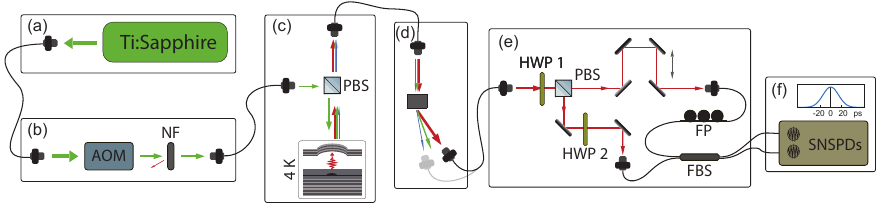}
    \caption{\textbf{Experimental set-up.} \textbf{(a)} A Ti:Sapphire laser produces laser pulses of 5 ps duration ($\lambda=923.238$~nm), with 90 GHz linewidth (green arrow). The spacing between laser pulses is 13.1 ns. \textbf{(b)} The excitation laser intensity is controlled with an acousto-optic modulator (AOM). Before the excitation of the QD, a narrow notch filter (bandwidth = 220 GHz full width at half maximum (FWHM)) ensures that any frequency components resonant with the transition being collected (here shown: XX) are removed from the excitation pulse. \textbf{(c)} The quantum dot, held at 4.2 K in a helium bath-cryostat, is excited to the $\ket{\text{XX}}$ via a two-photon excitation (TPE) process. The transmitted excitation laser undergoes a first filtering step using a cross-polarised microscope with a polarising beam splitter (PBS) to isolate the QD emission. In the schematic the micro-cavity is resonant with the XX (red arrows, $\lambda=924.223$~nm), and the collection of XX photons is strongly enhanced. A small fraction of the X emission (blue arrows, $\lambda=922.256$~nm) is also captured, compare Fig.~\ref{fig:spectra}(c). By in-situ tuning of the cavity length, the reverse case can be realised, i.e., enhancing and efficient collection of the X, and a small fraction collected from the XX emission. \textbf{(d)} A diffraction grating (bandwidth = 30~GHz FWHM) allows light from either the XX or X transition to be transmitted for further analysis. \textbf{(e)} Two-photon interference measurement setup. Half-wave plate 1 (HWP 1) and a polarising beam splitter (PBS) are used to control the ratio of light going into the two arms, to account for different losses in the two optical paths. The length of one arm can be adjusted to match the time-delay between the two arms to the spacing of the laser pulses. Half-wave plate 2 (HWP 2) and fibre paddles (FP) are used to change from the co-polarised to the cross-polarised configuration. The classical interference visibility is 98.5$\pm$1\%. The two paths are recombined in a fibre beam splitter (FBS), reflection:transmission = 0.525:0.475. \textbf{(f)} Photons are detected with superconducting nanowire single-photon detectors (SNSPDs). The combined timing jitter of the detectors and our photon counting electronics (Swabian Instruments Time Tagger Ultra) is 43~ps FWHM.}
    \label{fig:set-up}
\end{figure}

The set-up used for the experiments is shown in Fig.~\ref{fig:set-up}. See caption for details.

\section*{2. Purcell Enhanced Decay Rates}

\begin{figure}[t!]
    \centering
    \includegraphics{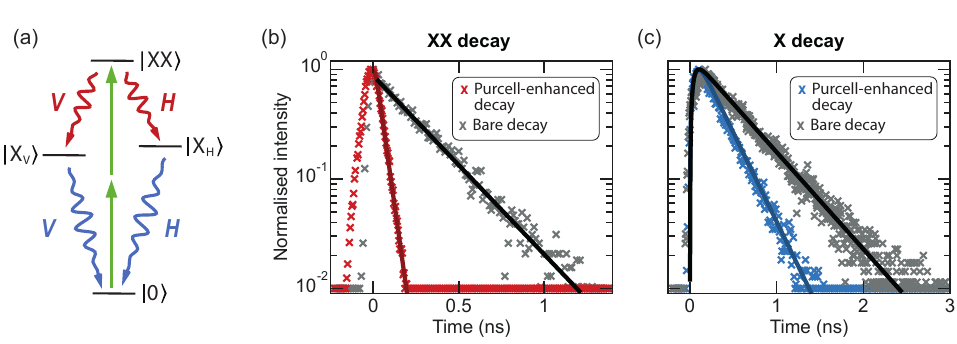}
    \caption{\textbf{Lifetime measurements following TPE, with and without cavity enhancement.} \textbf{(a)} Full level structure of the biexciton cascade. Green arrows: TPE. \textbf{(b)} Purcell-enhanced lifetime measurement of XX photons on resonance with the cavity (red) and bare lifetime with no Purcell enhancement (gray, measured with the cavity 690~GHz detuned from resonance). Solid lines: fit to exponential decay, convolved with a Gaussian instrument response function with FWHM 43~ps. \textbf{(c)} Purcell-enhanced lifetime measurement of X photons on resonance with the cavity (blue) and bare lifetime with no Purcell enhancement (gray, measured with the cavity 690~GHz detuned from resonance). Solid lines: Fit to Eq.~(\ref{eq:cascade_decay}) convolved with a Gaussian instrument response function with FWHM 43~ps.}
    \label{fig:lifetime_shortening}
\end{figure}

The bare lifetimes (measured with the microcavity 690~GHz detuned from the transition, enough to ensure no Purcell enhancement into the cavity mode [1]), are 263$\pm$7~ps for the $\ket{\text{XX}}$ and 484$\pm$6~ps for the $\ket{\text{X}}$ (error bars are one standard deviation determined from repeated measurements). This difference in lifetimes is expected, as the $\ket{\text{XX}}$ has two decay channels, while each $\ket{\text{XX}}$ has only one decay channel, Fig.~\ref{fig:lifetime_shortening}(a). Further, the dipole strength of the $\ket{\text{XX}}$ and the $\ket{\text{X}}$ can vary due to their different electronic environments [2]. The microcavity can then be tuned to introduce Purcell-enhancement on either the \decayXXtoX \ or the \decayXtoGS \ transition, resulting in minimum lifetimes of 38$\pm$1~ps and 43~ps$\pm$4 for the $\ket{\text{XX}}$ and $\ket{\text{X}}$, respectively (error bars are one standard deviation determined from repeated fitting using bootstrap resampling). This behavior is commonly quantified with the Purcell factor $F_\text{P}$, which for the transition $i$ is defined as $F_\text{P}^i = \frac{\gamma_i^c}{\gamma_i}$, where $\gamma_i^c$ is the decay rate with Purcell enhancement, and $\gamma_i$ is the bare decay rate. As stated in the main text, the Purcell factor for the $\ket{\text{XX}}$ (6.9) and the $\ket{\text{X}}$ (11.2) differ significantly in our experiments. This can be explained by taking into account the level structure of the biexciton cascade: the $\ket{\text{XX}}$ has two decay channels, the $\ket{\text{X}}$ only one. For the $\ket{\text{X}}$, the enhancement of the decay rate $f$ and the Purcell factor are identical: $F_\text{P}^X = \frac{\gamma_X^c}{\gamma_X} = \frac{f \gamma_X}{\gamma_X} = f = 11.3$. For the $\ket{\text{XX}}$ however, the situation is different. Since it has two decay channels, the total bare decay rate is given by twice the decay rate of a single decay channel $\gamma_{XX}^0$, i.e. $\gamma_{XX} = 2 \gamma_{XX}^0$. With a mode-split microcavity, only one of the two decay channels is enhanced, leading to an enhanced total decay rate of $\gamma_{XX}^c = \gamma_{XX}^0 + f \gamma_{XX}^0 = (1+f) \; \gamma_{XX}^0$. Therefore, $F_\text{P}^{XX} = \frac{\gamma_{XX}^c}{\gamma_{XX}} = \frac{1+f}{2}$. Assuming the same Purcell-enhancement $f$ for both $\ket{\text{XX}}$ and $\ket{\text{X}}$, we expect $F_\text{P}^{XX} \approx 6.1$, close to the measured value of 6.9. \\

Lifetime measurements with and without Purcell enhancement for the $\ket{\text{XX}}$ and $\ket{\text{X}}$ are shown in Fig.~\ref{fig:lifetime_shortening}(b,c). For the case where the $\ket{\text{XX}}$ ($\ket{\text{X}}$) is not Purcell-enhanced, the cavity is on the \decayXtoGS \ (\decayXXtoX) transition, 690 GHz detuned. The corresponding lifetimes were extracted by fitting the data. For the $\ket{\text{XX}}$, the fit is an exponential decay. For the $\ket{\text{X}}$, the cascade has to be taken into account. Using a rate equation approach yields 
\begin{equation}
    p_X(t) = p_{XX}^0 \left( \frac{\gamma_{_{XX}}}{\gamma_{_X} - \gamma_{_{XX}}} \right) \left[ \exp(-t \, \gamma_{_{XX}}) - \exp(-t \, \gamma_{_X}) \right],
    \label{eq:cascade_decay}
\end{equation}
where $p_X$ is the population of the $\ket{\text{X}}$, $p_{XX}^0$ is the initial population of the $\ket{\text{XX}}$, $\gamma_{_X}$ is the $\ket{\text{X}}$ decay rate and $\gamma_{_{XX}}$ is the $\ket{\text{XX}}$ decay rate. We note that the decay rate of the $\ket{\text{XX}}$ enters into the expression for the decay rate of the \decayXtoGS \  transition. $\ket{\text{XX}}$ lifetime measurements used for the fit in Fig.~\ref{fig:lifetime_shortening}(c) are not shown, but are comparable to data in Fig.~\ref{fig:lifetime_shortening}(b). All fitting functions were convolved with a Gaussian with FWHM 43~ps to account for the timing jitter of the detectors.

\section*{3. Spectra}

\begin{figure}[ht]
    \centering
    \includegraphics{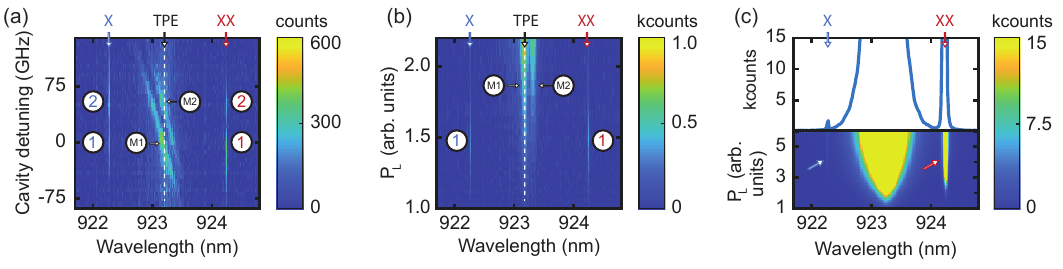}
    \caption{\textbf{Spectra under TPE.} \textbf{(a)} Scanning the cavity across the TPE frequency. When one of the two cavity modes is in resonance with the TPE frequency (M1, M2), the excitation is enhanced, leading to a strong emission of both $\ket{\text{X}}$ (\textcolor{c_X}{1},\textcolor{c_X}{2}) and $\ket{\text{XX}}$ (\textcolor{c_XX}{1},\textcolor{c_XX}{2}). The two cavity modes are split by 50 GHz and visible as bright, diagonal lines. The y-axis shows the detuning of the higher frequency cavity mode from the TPE. \textbf{(b)} Increasing the excitation laser power, with the higher frequency cavity mode in resonance with the TPE. The emission maxima of the $\ket{\text{X}}$ (\textcolor{c_X}{1}) and the $\ket{\text{XX}}$ (\textcolor{c_XX}{1}) occur at the same laser power (P$_\text{L}$ = 1.5), corresponding to a $\pi$-pulse on the TPE transition. The two cavity modes are visible as vertical lines (M1, M2). \textbf{(c)} Power series of the excitation laser, with the cavity in resonance with \decayXXtoX. The emission of the $\ket{\text{XX}}$, \textcolor{c_XX}{red arrow} ($\ket{\text{X}}$, \textcolor{c_X}{blue arrow}) is strong (weak). The top-part shows a line cut for the highest laser power.}
    \label{fig:spectra}
\end{figure}

Fig.~\ref{fig:spectra}(a) shows spectra on tuning the cavity across the TPE, Fig.~\ref{fig:spectra}(b) on fixing the cavity at the TPE and increasing the power of the excitation laser, and Fig.~\ref{fig:spectra}(c) on fixing the cavity on the \decayXXtoX \  transition and increasing the power of the excitation laser. For all cases, no frequency filtering (neither NF nor grating) was applied. In Fig.~\ref{fig:spectra}(a,b), the emission from the $\ket{\text{XX}}$ and the $\ket{\text{X}}$ is strongly correlated. The pair of bright, diagonal (a) or vertical (b) lines around the TPE resonance originate from the two cavity modes, due to either increased leakage of the excitation laser into the collection channel and/or direct cavity-enhanced two-photon emission $\ket{\rm{XX}}\rightarrow\ket{\rm{0}}$ [3-5]. Fig.~\ref{fig:spectra}(c) shows the very weak emission that is collected from the $\ket{\text{X}}$ when the cavity enhances \decayXXtoX. In the line cut in the upper part of (c), the frequency-overlap of the excitation laser with the two transitions is visible. For this reason, the NF in Fig.~\ref{fig:set-up}(b) is tuned to remove the frequency of the investigated transition from the excitation laser. 

\section*{4. Photon purity and coherence measurements}

\subsection*{4.1 \texorpdfstring{ g$^{(2)}$(0)}{g(2)(0)} measurements}
\begin{figure}[ht]
    \centering
    \includegraphics{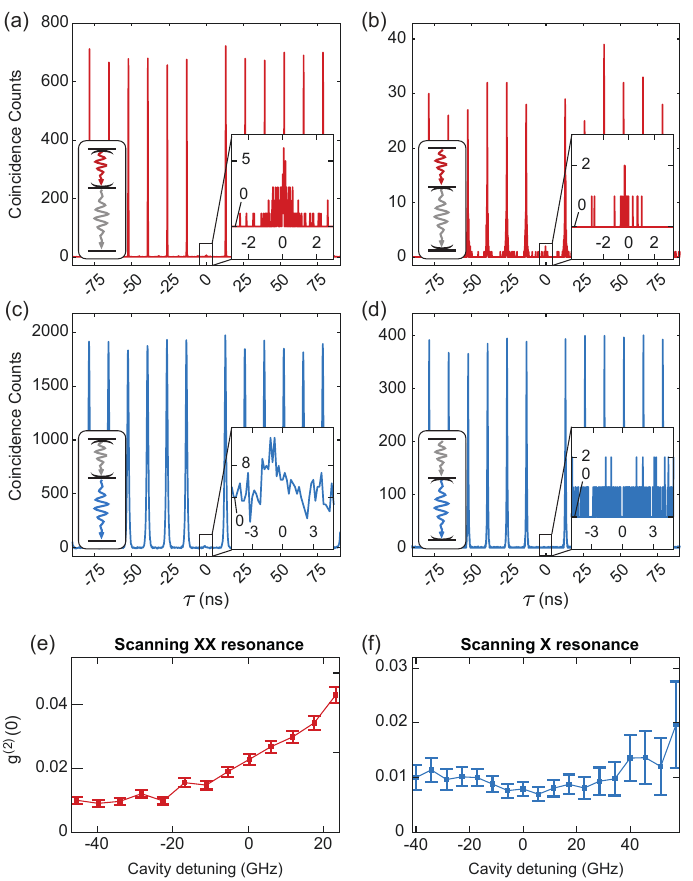}
    \caption{\textbf{g$^{(2)}(\tau)$ data for combinations of enhanced and collected transitions.} Coincidence counts for \textbf{(a)} enhancing and collecting XX, $g^{(2)}(0) = 2.3\pm0.2\%$, \textbf{(b)} enhancing X, collecting XX, $g^{(2)}(0) = 2.1\pm0.4\%$, \textbf{(c)} enhancing XX, collecting X, $g^{(2)}(0) = 1.1\pm0.1\%$, \textbf{(d)} enhancing and collecting X, $g^{(2)}(0) = 0.8\pm0.1\%$. \textbf{(e)} $g^{(2)}(0)$ values on scanning the cavity across the XX transition. \textbf{(f)} Same as (e) but for the X transition. All errors account for Poissonian statistics.}
    \label{fig:g2}
\end{figure}

The single photon purity is determined with the second order autocorrelation function at zero delay, g$^{(2)}$(0). Experimentally, the set-up shown in Fig.~\ref{fig:set-up}(e) with one of the two optical paths blocked is used to measure g$^{(2)}$(0) in a Hanbury Brown and Twiss experiment. Coincidences after pulsed TPE are shown in Fig.~\ref{fig:g2}(a-d) for each combination of enhanced and collected transitions. In all cases except enhancing and collecting XX, g$^{(2)}$(0) values were calculated by summing all counts in a 13.1~ns window around the central peak and normalising to the average counts in 10 side peaks, with the same integration window. In the case enhancing and collecting XX, a reflection in the system caused an additional signal peak centred at a delay around 6~ns. For this case, a 4~ns integration window was chosen (an argument why this is justified is given in subsection~4.3). In the case enhancing XX, collecting X, a significant number of background counts were recorded. For this case only, background subtraction was conducted as detailed in subsection~4.3. When either the XX or X transition is both enhanced and collected, g$^{(2)}$(0) measurements were conducted as a function of cavity detuning from the transitions. The resulting g$^{(2)}$(0) values lie between 0.7\% and 4.3\% and are shown in Fig.~\ref{fig:g2}(e,f). We attribute finite values for g$^{(2)}$(0) to imperfect excitation laser filtering and phonon-assisted cavity feeding. To improve the excitation laser suppression, a narrow NF (Fig.~\ref{fig:set-up}(b)) removes the frequency of the measured transition of the broad excitation pulse before it enters the cavity.

\subsection*{4.2 Photon coherence measurements}

Opening both paths in Fig.~\ref{fig:set-up}(e) allows us to measure the single photon coherence. Fig.~3 in the main text shows the coincidences detected on the two detector channels when the photons have the same polarisation (co-polarised configuration). For the experiments with high photon flux (enhancing and collecting the same transition), the cross-polarised configuration was also measured. HWP 2 and FP in Fig.~\ref{fig:set-up}(e) are used to adjust the relative polarisation of the photons in the two paths. For the high-flux experiments where both co-polarised and cross-polarised data was acquired, the HOM visibility is computed as \hbox{1 - $n_\parallel / n_\perp $}, where $n_\parallel$ ($n_\perp$) is the sum of the normalised counts in a 13.1~ns window around the central peak (thus covering a window significantly larger than the emitter lifetime and introducing no temporal post-selection). For both co- and cross-polarised data, the counts are normalised to the mean counts in 10 side peaks (using the same 13.1~ns integration window width). For the cases with low photon flux (enhancing and collecting different transitions), only the co-polarised case was measured, due to the long measurement times (32 h when collecting X, 49 h when collecting XX, compared to a few minutes for the high photon flux experiments). For these cases without cross-polarised data sets, the HOM visibility was calculated using the expression $1 - 2 n_\parallel$ [6]. On enhancing XX and collecting X, a significant number of background-counts was recorded. They were subtracted in the same way as for the g$^{(2)}$(0) measurement, detailed in subsection~4.3. On enhancing and collecting the same transition (high photon flux), we measured the photon coherence while scanning the cavity over the XX (X) transition: HOM values are shown in red (blue) in the inset to Fig.~4 in the main text. The values for the HOM visibility were corrected for the imperfect classical visibility of the set-up $1-\epsilon$ (98.5$\pm$1\%), finite values of g$^{(2)}$(0), and an imperfect 50:50 beam-splitter with reflection (R) : transmission (T) = 0.525$\pm$0.002 : 0.475$\pm$0.002, using the established correction [6] $V_{HOM} = (1-\epsilon)^{-2} (1+2g^{(2)}(0)) (R^2+T^2)/(2RT) V_{raw}$.

\subsection*{4.3 Background corrections}

\begin{figure}
    \centering
    \includegraphics[scale=0.98]{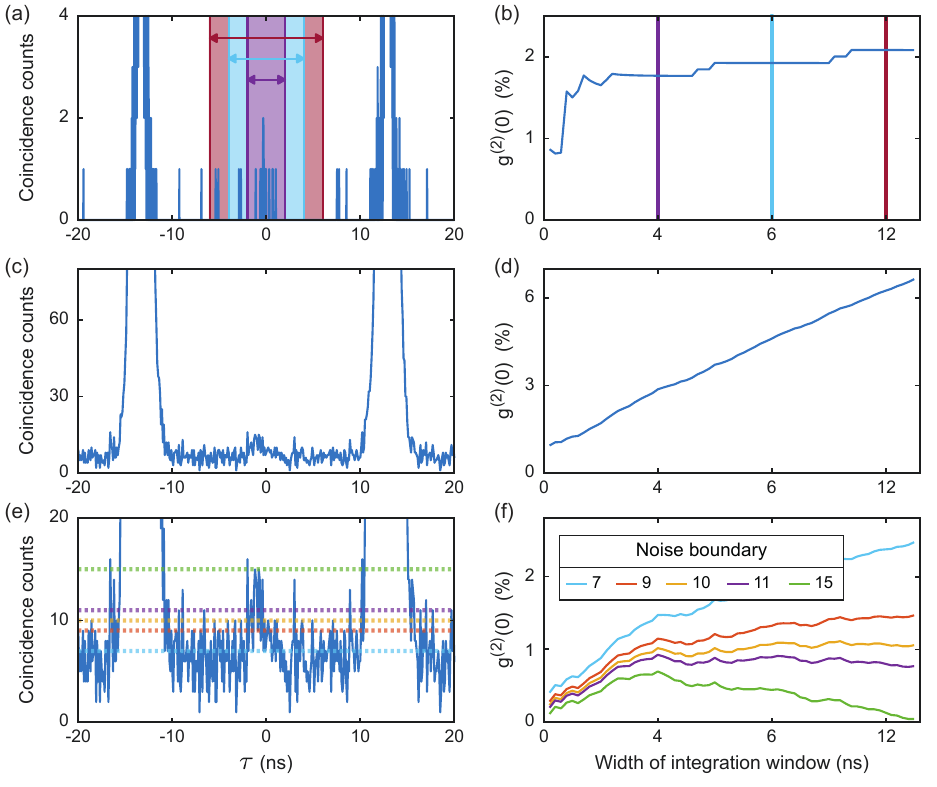}
    \caption{\textbf{Background subtraction.} \textbf{(a)} Zoom-in of g$^{(2)}(\tau)$ data from Fig.~\ref{fig:g2}(b), a measurement with relatively low background signal. \textbf{(b)} g$^{(2)}$(0) values on changing the width of the integration window in data from (a). Purple, blue, red lines correspond to integration windows shown in (a). \textbf{(c)} Zoom-in of g$^{(2)}(\tau)$ data from Fig.~\ref{fig:g2}(c), a measurement with high background signal. \textbf{(d)} g$^{(2)}$(0) values when changing the width of the integration window in data from (c). \textbf{(e)} Further zoom-in of (c), with horizontal, dashed lines depicting possible noise boundaries (see legend in (f)). \textbf{(f)} g$^{(2)}$(0) values on changing the width of the integration window in data from (e), with different noise boundaries. For the curve for noise boundary = 10, g$^{(2)}$(0) reaches a plateau, identical to the behavior for a noiseless measurement. Therefore, this is the correct value for the background subtraction.}
    \label{fig:bg_sub}
\end{figure}

On enhancing XX and collecting X (low photon-flux experiments), the signal-to-noise ratio was considerably lower and during the long integration time, a significant number of background counts was recorded for both g$^{(2)}$(0) and HOM visibility measurements. They originate from background lights and have no temporal correlations. This background signal would create systematic errors in the extracted numbers for g$^{(2)}$(0) and HOM visibility, and was therefore subtracted in the following way (explanation for g$^{(2)}$(0), identical process for HOM visibility). In an ideal case without any background signal, increasing the integration window of the coincidence counts does not affect values for g$^{(2)}$(0) after a certain point (typically around 2~ns), Fig.~\ref{fig:bg_sub}(a,b). However, with the background signal, an increase in integration window decreases the apparent photon purity, as more counts are collected in the 0-delay window, Fig.~\ref{fig:bg_sub}(c,d). Therefore, the background signal is subtracted, so that g$^{(2)}$ vs. integration window length plateaus again. The proper background level is found by taking the average coincidence counts which are smaller than a certain noise boundary. To find the correct noise boundary, the behavior of g$^{(2)}$(0) versus integration window width for different noise boundaries is computed, Fig.~\ref{fig:bg_sub}(e,f). For the correct value, g$^{(2)}$(0) plateaus, retrieving the behavior of a noiseless measurement, which is the case here for a noise boundary of 10 coincidence counts.

\newpage
\begin{enumerate}[label={[\arabic*]}, nosep]
\item M.~R. Hogg, N.~O. Antoniadis, M.~A. Marczak, G.~N. Nguyen,
T.~L. Baltisberger, A. Javadi, R. Scholl, S.~R. Valentin,
A.~D. Wieck, A. Ludwig \& R.~J. Warburton.
Fast optical control of a coherent hole spin in a micocavity.
\href{https://www.nature.com/articles/s41567-025-02988-5}{Nature Physics \textbf{21}, 1475 (2025)}.

\item P.~A. Dalgarno, J.~M. Smith, B.~D. McFarlane, B.~D. Gerardot,
K. Karrai, A. Badolato, P.~M. Petroff \& R.~J. Warburton.
Coulomb interactions in single charged self-assembled quantum dots:
Radiative lifetime and recombination energy.
\href{https://journals.aps.org/prb/abstract/10.1103/PhysRevB.77.245311}{Physical Review B \textbf{77}, 245311 (2008)}.

\item S. Schumacher, J. F\"orster, A. Zrenner, M. Florian, C. Gies,
P. Gartner \& F. Jahnke.
Cavity-assisted emission of polarization-entangled photons from
biexcitons in quantum dots with fine-structure splitting.
\href{https://opg.optica.org/oe/fulltext.cfm?uri=oe-20-5-5335}{Optics Express \textbf{20}, 5335 (2012)}.

\item D. Heinze, A. Zrenner \& S. Schumacher.
Polarization-entangled twin photons from two-photon quantum-dot emission.
\href{https://journals.aps.org/prb/abstract/10.1103/PhysRevB.95.245306}{Physical Review B \textbf{95}, 245306 (2017)}.

\item S. Liu, Y. Wang, Y. Saleem, X. Li, H. Liu, C.-A. Yang, J. Yang,
H. Ni, Z. Niu, Y. Meng, X. Hu, Y.~Y. Xuehua Wang, M. Cygorek
\& J. Liu.
Quantum correlations of spontaneous two-photon emission from a quantum dot.
\href{https://www.nature.com/articles/s41586-025-09267-6}{Nature \textbf{643}, 1234 (2025)}.

\item C. Santori, D. Fattal, J. Vuckovic, G.~S. Solomon \& Y. Yamamoto.
Indistinguishable photons from a single-photon device.
\href{https://www.nature.com/articles/nature01086}{Nature \textbf{419}, 594 (2002)}.
\end{enumerate}

\end{document}